\documentclass[12pt]{article}
\usepackage[T1]{fontenc}
\usepackage[utf8]{inputenc}
\usepackage{authblk}
\usepackage{cite}
\usepackage[dvipsnames]{xcolor}
\usepackage{graphicx,graphics}
\usepackage{caption,subcaption}
\begin{document}
\title{Numerical Solutions in 5D Standing Wave Braneworld}
\author[1,2]{Merab Gogberashvili \thanks{gogber@gmail.com}}
\author[1]{Otari Sakhelashvili \thanks{otosaxel@gmail.com}}
\author[1]{Giorgi Tukhashvili \thanks{gtukhashvili10@gmail.com}}
\affil[1]{I. Javakhishvili Tbilisi State University, 3 Chavchavadze Avenue,

Tbilisi 0179, Georgia}

\affil[2]{E. Andronikashvili Institute of Physics, 6 Tamarashvili Street,

Tbilisi 0177, Georgia}

\maketitle

\begin{abstract}
Within the 5D standing wave braneworld model numerical solutions of the equations for matter fields with various spins are found. It is shown that corresponding action integrals are factorizable and convergent over the extra coordinate, i.e. 4D fields are localized on the brane. We find that only left massless fermions are localized on the brane, while the right fermions are localized in the bulk. We demonstrate also quantization of Kaluza-Klein excited modes in our model.


\end{abstract}

\section{Introduction}

Braneworld models with large extra dimensions \cite{Hi-1,Hi-2,brane-1,brane-2} have attracted wide attention owing to their success in solving some open questions in modern physics \cite{reviews-1,reviews-2,reviews-3,reviews-4}. To realize the braneworld idea one should find a mechanism of localization, however, there are some difficulties in this matter within standard brane approaches \cite{BaGa,Od}. It is especially economical to consider models with a pure gravitational localization mechanism, since gravity has universal couplings with all matter fields. In this paper we consider one such scenario within the standing waves braneworld \cite{Wave-1,Wave-2} and explicitly show existence of the universal gravitational trapping of all kinds of matter fields on the brane. Main differences of the model with the most brane scenarios are that our metric {\it ansatz} is time-dependent and contains increasing warp factor.

The standing waves braneworld is formed by the background solution \cite{Wave-1,Wave-2}:
\begin{equation} \label{metric}
ds^2 = e^{2a|r|}\left( dt^2 - e^{u}dx^2 - e^{u}dy^2 - e^{-2u}dz^2 \right) - dr^2~,
\end{equation}
of the coupled Einstein and massless ghost scalar field equations. In (\ref{metric}) $r$ denotes the extra space-like coordinate, $a>0$ is a curvature scalar (we choice increasing warp factor), the determinant,
\begin{equation} \label{determinant}
\sqrt g = e^{4a|r|}~,
\end{equation}
is $u$-independent and the oscillatory function $u(t,r)$ has the form:
\begin{eqnarray} \label{u}
u(t,r) = \sin (\omega t) Z(r)~, \nonumber \\
Z(r) = C e^{-2a|r|} Y_2\left( \frac{\omega}{a} e^{-a|r|} \right)~.
\end{eqnarray}
Here $\omega$ denotes the oscillation frequency, $C$ is an integration constant and $Y_2(r)$ is the second-order Bessel function of the second kind. The solution (\ref{metric}) describes the brane located at the node of the standing bulk wave (\ref{u}). In this paper we consider the case with one node, i.e. the function $Y_2(r)$ in (\ref{u}) has a single zero at the position of the brane, $r = 0$. This is accomplished by the assumption:
\begin{equation}\label{FirstZerosY}
\frac{\omega}{a} \approx 3.38~.
\end{equation}

In the equations of the matter fields, the oscillatory function (\ref{u}) enters via exponentials. We suppose that the frequency $\omega$ of standing waves is much larger than the frequencies associated with the energies of particles on the brane and in these equations we can perform the time averaging of oscillating exponents. Non-vanishing averages are expressed by the formula \cite{JHEP,IJMP,GMM-1}:
\begin{equation} \label{e-u}
\left\langle e^{u} \right\rangle = I_0 \left( Z \right) ~,
\end{equation}
where $I_0(r)$ is the modified Bessel function.

The brane at the node of the standing wave (the point where $Z (r)$ and $u(t,r)$ vanish) can be considered as a 4D 'island', where the matter particles are assumed to be bound. Indeed the system of 5D geodesic equations of motion for a classical particle (or a photon) reduces to \cite{JHEP}:
\begin{equation} \label{dr}
\frac 12 e^{-2a|r|} \left(\frac{d r}{d t}\right)^2 + a(1-V^2) |r| = \epsilon~,
\end{equation}
where $V$ is the velocity along the brane and the constant $\epsilon$ corresponds to the energy of the particle per unit mass (for massive particles $ 0< \epsilon = E/m < 1$ and for photons $\epsilon = 1$). From (\ref{dr}) it is clear that the motion towards the extra dimension $r$ is possible when
\begin{equation}
\epsilon - a (1-V^2) |r| \geq 0~.
\end{equation}
The term $a(1-V^2) |r|$ plays the role of the trapping gravitational potential and there exists maximal distance in the bulk $|r|_{max} \sim \epsilon/a$. In the standard brane approach with decreasing warp factor ($a<0$) localization was achieved due to the fact that the extra space actually is finite \cite{brane-1,brane-2}. In our case, the increasing of the brane warp factor, $e^{2a|r|}$, creates the potential well that confines particles.

Let us now consider the localization problem of quantum fields with various spin using the brane solution (\ref{metric}) \cite{JHEP,IJMP,GMM-1, GMM-2}. To have localized fields on a brane 'coupling' constants appearing after integration of their Lagrangians over the extra coordinate, $r$, must be non-vanishing and finite. Using numerical solutions we shall demonstrate existence of normalizable zero modes of various fields on the brane within the model.


\section{Localization of bosons}

In this section, using numerical solutions, it is shown localization of massless modes of the spin-0, spin-1 and spin-2 particles on the brane.

At first let us consider 5D massless scalar field. We assume that for the case $\omega \gg E$ ($E$ is the typical energy of scalar particles on the brane) variables can be separated \cite{JHEP,IJMP,GMM-1},
\begin{equation}\label{Solution1}
\Phi \left( x^A \right) =\phi (x^\nu) \varsigma (r)~,
\end{equation}
Capital Latin letters numerate 5D coordinates, while Greek ones are used for 4D indices. Then the action integral for scalar fields can be split in two parts:
\begin{equation} \label{Sphi}
S_\Phi = - \frac 12 \int d^5x\sqrt{g} ~ g^{MN}\partial_M\Phi \partial_N\Phi = - \frac 12 \int d^4x \left[\partial_\alpha \phi \partial^\alpha \phi\int dr e^{2a|r|}\varsigma^2 - \phi^2\int dr e^{4a|r|}\varsigma'^2\right]~,
\end{equation}
where prime denotes the derivative with respect to extra coordinate $r$. We are looking for zero mode solution on the brane, $\phi \sim e^{ip_\nu x^\nu}$, and assume that energy-momentum along the brane, $p_\nu$, obeys the dispersion relation:
\begin{equation} \label{dispersion}
E^2 - p_x^2 - p_y^2 - p_z^2 = 0~.
\end{equation}
Than the 5D Klein-Gordon equation after time averaging reduces to \cite{JHEP,IJMP,GMM-1}:
\begin{equation}\label{xi-2}
\left( e^{4a|r|}\varsigma' \right)' - e^{2a|r|}P^2(r) \varsigma = 0~,
\end{equation}
where
\begin{equation} \label{P2}
P^2(r) = \left(\left\langle e^{-u} \right\rangle -1\right)\left( p_x^2 + p_y^2 \right) + \left(\left\langle e^{2u} \right\rangle -1\right)p_z^2 ~.
\end{equation}


\begin{figure}[ht]
\begin{center}
\includegraphics[width=0.7\textwidth]{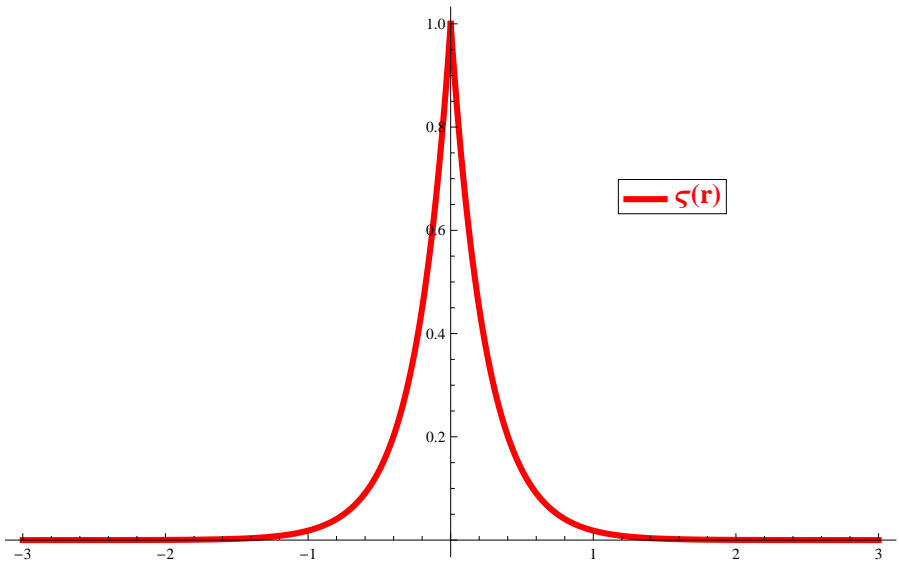}
\caption{Scalar zero mode's profile in the bulk.}
\end{center}
\end{figure}


Close to the brane the function $P^2(r)$, which describes oscillatory properties of standing waves, and the extra dimension part of wave function $\varsigma(r)$, behave as:
\begin{eqnarray} \label{solution-s}
\left. P^2 (r)\right|_{r \to 0} &\sim& r^2~, \nonumber \\
\left. \varsigma (r)\right|_{r \to 0} &\sim& e^{-4a|r|}~.
\end{eqnarray}
So for the equation (\ref{xi-2}) we use the boundary conditions:
\begin{equation}
\varsigma(0) = 1~, ~~~~~ \varsigma'(0) = - 4a~.
\end{equation}
Figure 1 shows the numerical solution of the equation (\ref{xi-2}) for these boundary conditions for the following values of the parameters:
\begin{equation}
a=1~, ~~~~~ \omega  =3.38 \sim a~, ~~~~~ p = 0.01 \ll a~.
\end{equation}
We see that $\varsigma (r)$, which governs of probability of scalar particles to reach the bulk, falls off from the brane to zero.

The integrals over $r$ in the action (\ref{Sphi}) are convergent if integrand functions decrease stronger than $1/r$. Figure 2 displays the products of $r$ on these integrand functions. We see that they decrease, i.e. the integrals over $r$ in the action (\ref{Sphi}) are convergent, hence 4D scalar fields are localized on the brane.


\begin{figure}[ht]
\begin{center}
\includegraphics[width=0.7\textwidth]{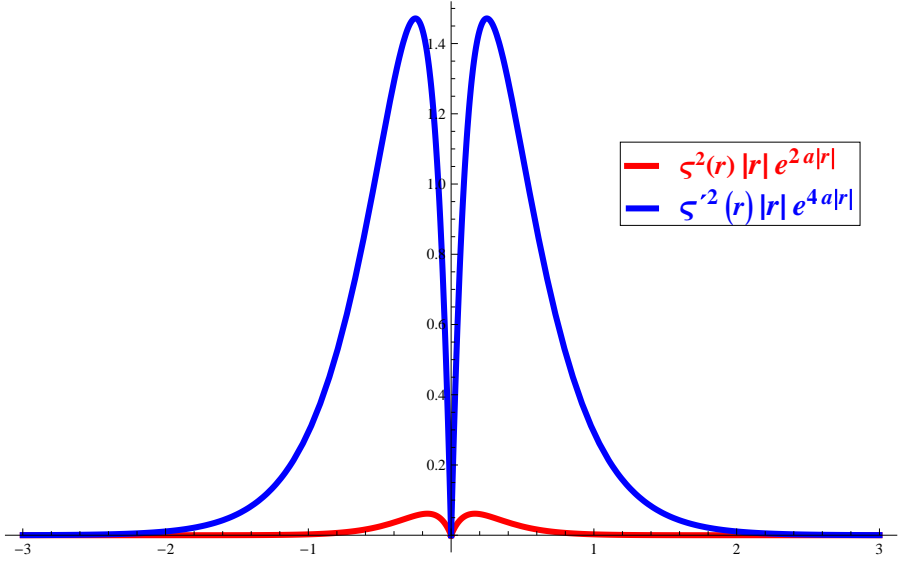}
\caption{Convergence of integrand functions in $S_\Phi$ (\ref{Sphi}).}
\end{center}
\end{figure}


Recall that the action of scalar fields (\ref{Sphi}) due to the determinant (\ref{determinant}) contains increasing exponential factor ($a > 0$). This is the reason why in the original brane models \cite{brane-1,brane-2} the scalar field zero modes with the constant extra dimension parts can be localized on the brane only in the case of decreasing warp factor (i.e. $a<0$). In our model the extra part of wave function (\ref{solution-s}) is not constant, moreover, from Figure 1 it is clear that it strongly decreases and according to Figure 2 the integrals over $r$ in the action (\ref{Sphi}) are convergent.

It is known that the transverse traceless graviton modes obey the equations similar to the massless scalar fields in a curved background. Accordingly, the condition of localization of spin-$2$ graviton field is equivalent to that of scalar field considered above \cite{JHEP,IJMP}.


\begin{figure}[ht]
\begin{center}
\includegraphics[width=0.7\textwidth]{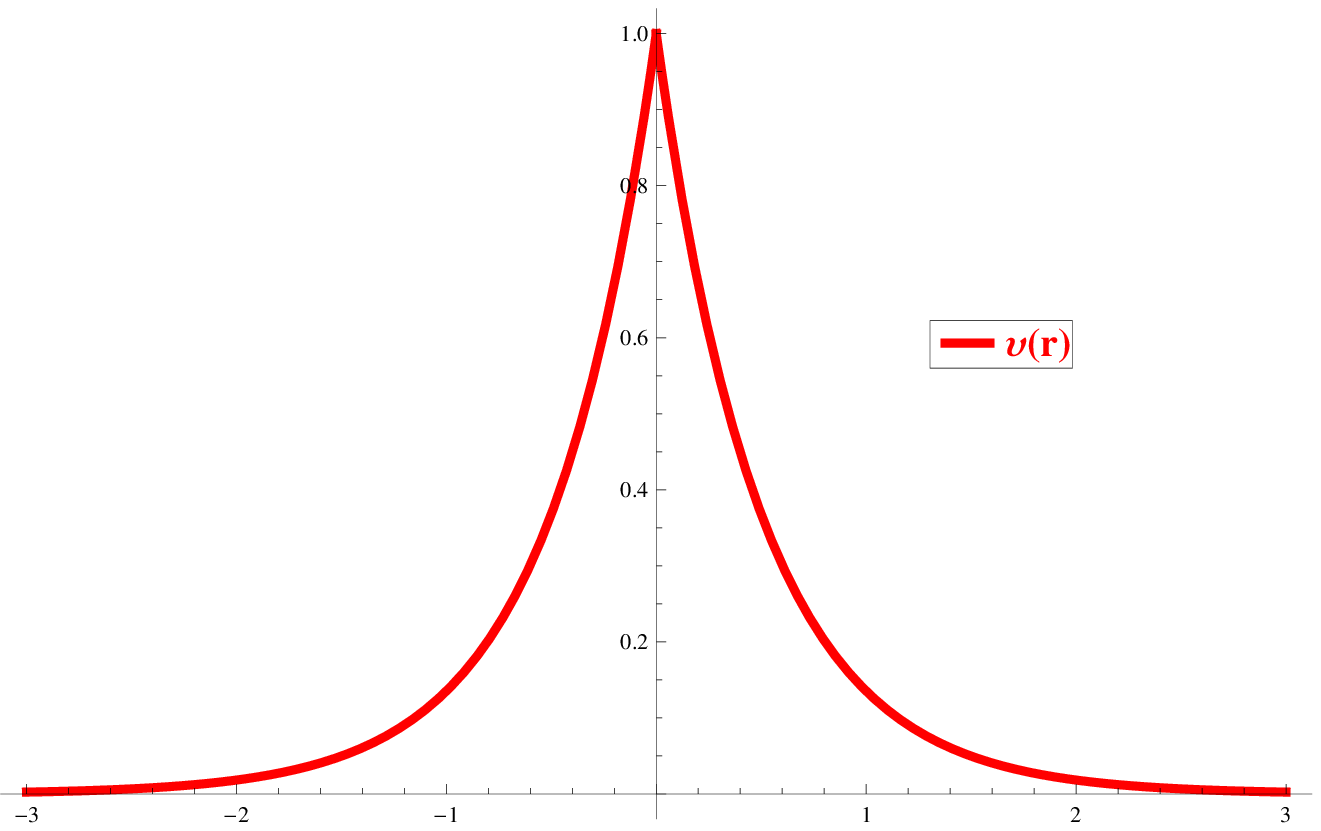}
\caption{Vector zero mode's profile in the bulk.}
\end{center}
\end{figure}


In the case of 5D Abelian vector fields (the generalization to the case of non-Abelian gauge fields is straightforward) we are looking for the solution of field equations in the form \cite{JHEP,IJMP,GMM-2}:
\begin{equation} \label{A}
{\cal A}_\alpha (x^A)= e^{-2a|r|} g_{\alpha\beta} A^\beta (x^\nu) v(r)~, ~~~~~ {\cal A}_r (x^A) = 0~.
\end{equation}
Than the 5D action of vector fields can be written as:
\begin{eqnarray}\label{VectorAction}
S_A = - \frac14\int d^5x\sqrt g~ g^{MN}g^{PR}{\cal F}_{MP}{\cal F}_{NR} = \nonumber \\
= - \frac14\int d^4x \left[ F_{\alpha\beta}F^{\alpha\beta}\int dr v^2 - 2A_\alpha A^\alpha \int dr e^{2a|r|}v'^2\right] ~.
\end{eqnarray}
We require the existence of 4D vector zero mode on the brane,
\begin{equation} \label{Factors}
A_\mu\left(x^\nu\right) \sim \varepsilon_\mu e^{ip_\nu x^\nu} ~,
\end{equation}
where $p_\nu$ obey dispersion relation (\ref{dispersion}). Then 5D Maxwell equations after time averaging yields the single equation for $v (r)$:
\begin{equation}\label{v}
\left( e^{2a|r|}v' \right)' - P^2(r) v  = 0~,
\end{equation}
where $P^2(r)$ is done by (\ref{P2}). Close to the brane, $\left. v (r)\right|_{r \to 0} \sim e^{-2a|r|}$, and we set the boundary conditions:
\begin{equation}
v (0) = 1~, ~~~~~ v'(0) = - 2a~.
\end{equation}
From Figure 3 we see that the solution of the equation (\ref{v}) for these boundary conditions, or the probability of photon to leave the brane, falls off in the bulk down to zero.

As for the case of scalar field the integrals over $r$ in the vector functions action (\ref{VectorAction}) are convergent if integrand functions decrease stronger than $1/r$. From the Figure 4 we see that the products of $r$ with integrand functions decrease, i.e. the integral over $r$ in (\ref{VectorAction}) is convergent and 4D photon is localized on the brane.


\begin{figure}[ht]
\begin{center}
\includegraphics[width=0.7\textwidth]{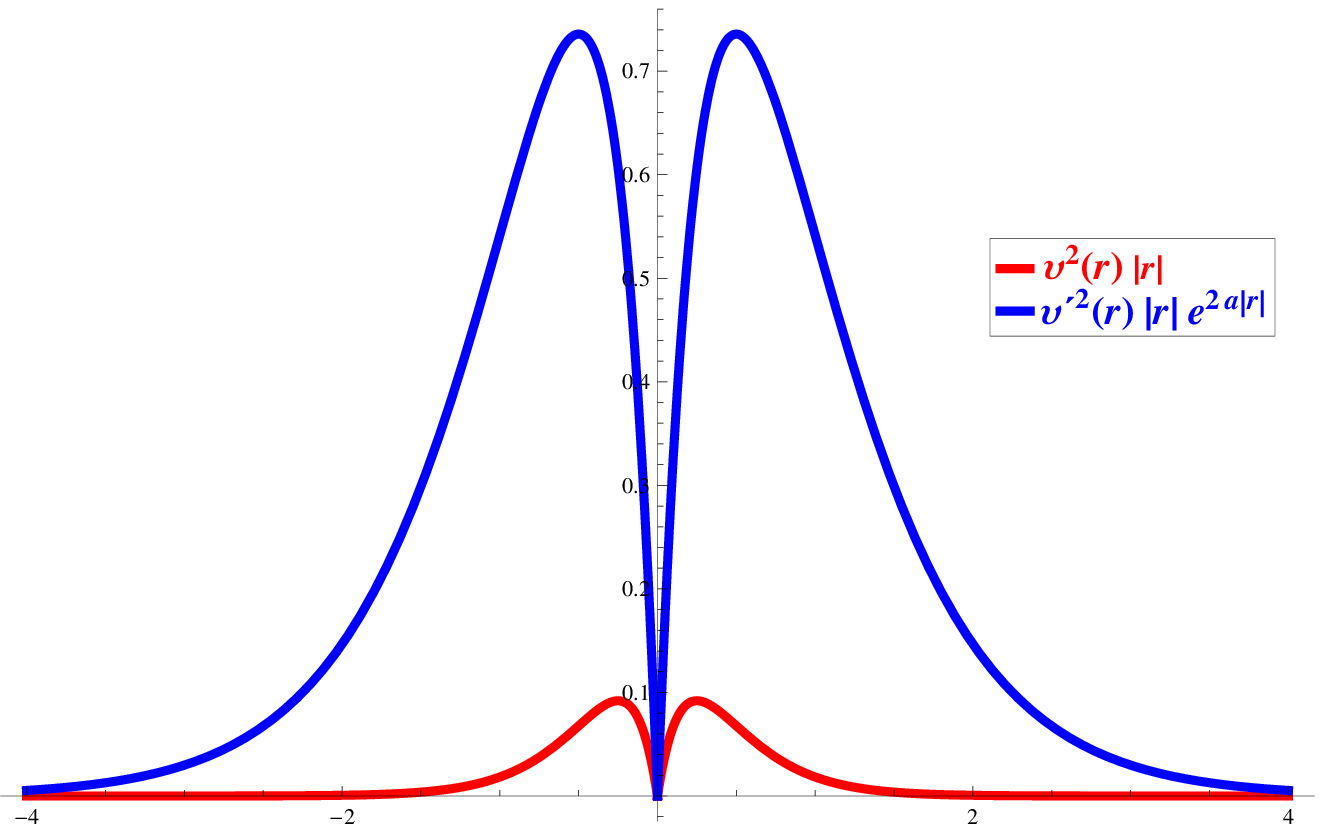}
\caption{Convergence of integrand functions in the vector field action (\ref{VectorAction}).}
\end{center}
\end{figure}


In 5D brane models the implementation of pure gravitational trapping mechanism of vector field particles remains the most problematic \cite{BaGa,Od}. The reason is that in the vector action (\ref{VectorAction}) the extra dimension parts of the determinant (\ref{determinant}) and two metric tensors with upper indices cancel each other. Because of this in the original brane models the vector field zero modes (with the constant extra dimension part) cannot be localized on the brane for any sign of $a$. In our model the extra dimension part of vector field $v (r)$ is rapidly decreasing function (Fig. 3) and the integral over $r$ in (\ref{VectorAction}) is convergent.


\section{Localization of fermions}

Now we investigate the localization problem for massless fermions \cite{JHEP,IJMP}.

For Minkowskian $4\times 4$ gamma matrices we use the Weyl basis,
\begin{equation}\label{MinkowskianGammaMatrices}
\begin{array}{l}
\gamma ^t =~ \left( {\begin{array}{*{20}{c}}
0&I\\
I&0
\end{array}} \right),~~~
{\gamma ^i} = \left( {\begin{array}{*{20}{c}}
0&-\sigma^i\\
\sigma^i&0
\end{array}} \right),~~~
\gamma ^5 = i \gamma^t\gamma^x\gamma^y\gamma^z = \left( {\begin{array}{*{20}{c}}
I&0\\
0&-I
\end{array}} \right),
\end{array}
\end{equation}
and 5D gamma matrices $\Upsilon^A = h_{\bar A}^A\Upsilon^{\bar A}$, where $h_{\bar A}^A = (e^{ - a|r|},e^{ - a|r| - u/2}, e^{ - a|r| - u/2}, e^{ - a|r| + u}, 1)$. The inverse {\it f\"unfbein} for the metric (\ref{metric}), we choice as:
\begin{equation}\label{GammaMatricesRelation}
\Upsilon ^\nu = h_{\bar\mu}^\nu ~\gamma ^\mu ~, ~~~~~\Upsilon ^r = i\gamma ^5~. \nonumber
\end{equation}

Non-vanishing components of the spin-connection in the background (\ref{metric}), which enter covariant derivatives defined as:
\begin{equation}
D_A = \partial_A + \frac 14 \Omega_A^{\bar B \bar C} \Upsilon_{\bar B} \Upsilon_{\bar C}~,
\end{equation}
after time averaging are given by:
\begin{equation}\label{Spin-ConnectionComponents}
\left\langle \Omega_\mu^{\bar \nu \bar r} \right\rangle = a ~sgn(r) \left\langle h_\mu^{\bar \nu}\right\rangle ~.
\end{equation}

For the wavefunction of the bulk fermion field we use the chiral decomposition:
\begin{equation}\label{Psi}
\Psi \left(x^\nu,r\right) = \psi_L \left(x^\nu\right) \lambda (r) + \psi_R \left(x^\nu\right) \rho (r)~,
\end{equation}
where $\lambda(r)$ and $\rho(r)$ are the extra dimension factors of the left and right fermion wavefunctions respectively.

Then the 5D Dirac action for massless fermions can be written as:
\begin{eqnarray}\label{SpinorAction}
S_\Psi = \int d^5x \sqrt g \left(\overline \Psi  i\Upsilon ^MD_M \Psi + H.c. \right) = \nonumber \\
= \int d^4x \left\{\overline \psi_L  i\gamma^\mu \partial_\mu \psi_L \int dr e^{3a|r|}\lambda^2 + \overline \psi_R  i\gamma^\mu \partial_\mu \psi_R \int dr e^{3a|r|}\rho^2 +\right. \\
\left. + \overline \psi_R \psi_L \int dr e^{4a|r|}\rho \left[\lambda' + 2a~ sgn(r) \lambda \right] - \overline \psi_L \psi_R \int dr e^{4a|r|}\lambda \left[\rho' + 2a~ sgn(r) \rho \right]\right\}~.\nonumber
\end{eqnarray}

We assume that 4D left and right Dirac spinors ($\gamma^5 \psi_{R/L} = \pm \psi_{R/L}$) are solutions of the free 4D massless Dirac equations:
\begin{eqnarray} \label{psi-free}
\psi_R (x^\nu) = \left( \begin{array}{c}R\\0\end{array} \right) e^{-ip_\nu x^\nu}, \nonumber \\
\psi_L (x^\nu) = \left( \begin{array}{c}0\\L\end{array} \right) e^{-ip_\nu x^\nu},
\end{eqnarray}
where $p_\nu$ satisfy the dispersion relation (\ref{dispersion}) and the constant 2-spinors $L$ and $R$ obey:
\begin{equation} \label{relations}
\left(p_t + \sigma^ip_i\right)L = \left(p_t - \sigma^ip_i\right)R = 0~.
\end{equation}

Derived from (\ref{SpinorAction}) 5D Dirac equation, after time averaging, takes the form:
\begin{equation}\label{SpinorEquation}
ie^{-a|r|}\left[\gamma^t\partial _t + \left\langle e^{ u/2} \right\rangle \left(\gamma ^x\partial _x + \gamma ^y\partial _y\right) + \left\langle e^{-u} \right\rangle \gamma ^z\partial _z\right]\Psi -\gamma^5 \left[2a ~sgn(r) + \partial_r \right]\Psi =0~.
\end{equation}
Using the expressions (\ref{psi-free}) and (\ref{relations}) it can be rewritten as the system:
\begin{eqnarray}\label{L-R}
\left( \begin{array}{*{20}{c}}
-e^{a|r|}\left[2a ~sgn(r) + \partial_r \right]&\sigma^i{\cal P}_i(r)\\
- \sigma^i{\cal P}_i(r)&e^{a|r|}\left[2a ~sgn(r) + \partial_r \right]
\end{array} \right)\left( \begin{array}{*{10}{c}} \rho (r)R\\
\lambda (r)L\end{array} \right) = 0~.
\end{eqnarray}
Here we have introduced the functions ${\cal P}_i (r)$:
\begin{eqnarray} \label{P-i}
{\cal P}_x (r) &=& \left(\left\langle e^{-u/2} \right\rangle -1\right) p_x = \left[I_0\left(Z/2\right)-1\right] p_x, \nonumber \\
{\cal P}_y (r) &=& \left(\left\langle e^{-u/2} \right\rangle - 1 \right)p_y = \left[I_0\left(Z/2\right)-1\right] p_y, \nonumber \\
{\cal P}_z (r) &=& \left(\left\langle e^{u} \right\rangle - 1 \right)p_z = \left[I_0\left(Z\right)-1\right] p_z,
\end{eqnarray}
where $Z(r)$ is defined in (\ref{u}).


\begin{figure}[ht]
\begin{center}
\includegraphics[width=0.7\textwidth]{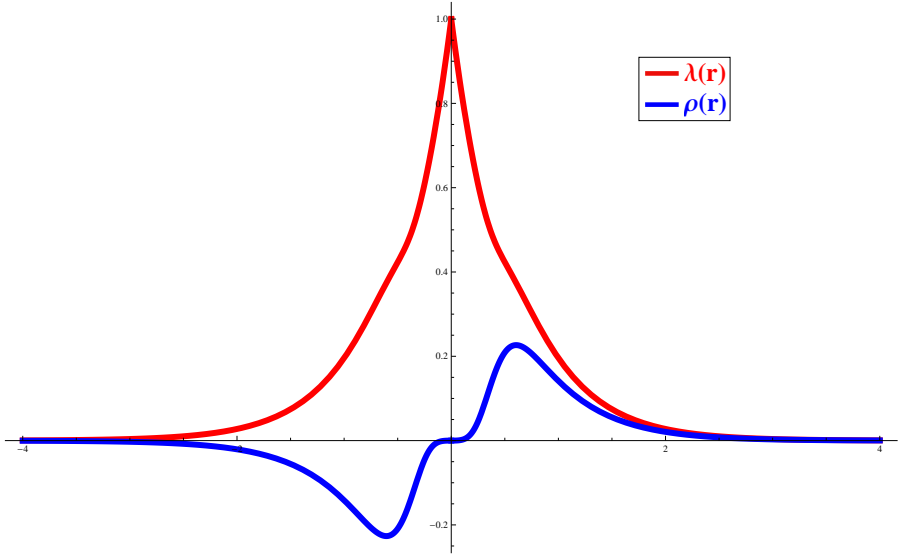}
\caption{Profiles of the left and right fermion wavefunctions in the bulk.}
\end{center}
\end{figure}


To find boundary conditions for the system (\ref{L-R}) note that from the last two terms of (\ref{SpinorAction}) we see that the action is even under the $Z_2$ orbifold symmetry if $\lambda (r) $ is an even function of $r$ and $\rho (r) $ is odd (or vice versa) with respect to the transformation $r \leftrightarrow -r$. We require,
\begin{equation}
\lambda_k(- r) = \lambda_k(r)~, ~~~~~ \rho_k(- r) = - \rho_k(r)~,
\end{equation}
what can be accomplished if we impose that the bulk fermion wavefunction is even under five-dimensional parity \cite{chiral-1,chiral-2}:
\begin{equation} \label{gammaPsi}
\gamma^5\Psi \left(x^\alpha, -r\right) = - \Psi \left(x^\alpha, r\right)~.
\end{equation}
Then from the symmetry of the problem with respect to the transformation $r \leftrightarrow -r$ we get the following important boundary conditions:
\begin{eqnarray} \label{baundary}
\rho (0) = 0~,~~~~~\lambda (0) = 1~.
\end{eqnarray}

Solutions of the Dirac equations (\ref{L-R}) with the boundary conditions (\ref{baundary}) are displayed on the Figure 5. We see that in our model the extra dimension part of the left spinor wave function $\lambda(r)$ has maximum on the brane and decreases in the bulk. While, as a consequence of (\ref{baundary}), in our setup the right fermionic modes are absent on the brane and $\rho (r)$ has maximum in the bulk outside the brane.

Integrals over $r$ in the spinor field action (\ref{SpinorAction}) are convergent if integrand functions decrease stronger than $1/r$. This feature for (\ref{SpinorAction}) is demonstrated on the Figure 6. So in our model left massless fermions are localized on the brane and right fermions are localized in the bulk.


\begin{figure}[ht]
\begin{center}
\includegraphics[width=0.7\textwidth]{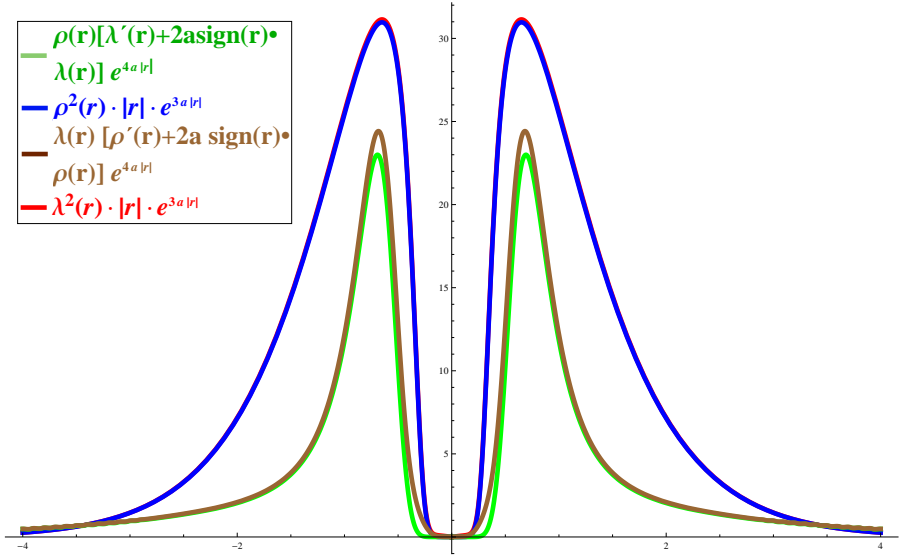}
\caption{Convergence of bulk integrals in $S_\Psi$ (\ref{SpinorAction}).}
\end{center}
\end{figure}


\section{Massive modes}

In this section we want to estimate masses of Kaluza-Klein excitations.

For simplicity let us consider the case of 5D scalar particles. For a massive mode the dispersion relation (\ref{dispersion}) takes the form:
\begin{equation} \label{dispersion-m}
E^2 - p_x^2 - p_y^2 - p_z^2 = m^2~,
\end{equation}
where $m$ denotes the mass of lightest KK mode on the brane. Then the equation for scalar particles (\ref{xi-2}) transforms to:
\begin{equation}\label{varsigma-m}
\left( e^{4a|r|}\varsigma' \right)' - e^{2a|r|}\left[P^2(r)-m^2 \right]\varsigma = 0~.
\end{equation}
It is more convenient to put (\ref{varsigma-m}) into the form of an analogue non-relativistic quantum mechanical problem by making the change:
\begin{equation} \label{varsigma}
\varsigma (r) = e^{-2a|r|} \psi (r)~.
\end{equation}
For $\psi (r)$ we find:
\begin{equation}\label{psi}
\psi'' +e^{-2a|r|} \left[ m^2- V(r) \right] \psi = 0~,
\end{equation}
where the function,
\begin{equation}\label{U}
V (r)= 4ae^{-2a|r|} \delta (r) + 4a^2 e^{-2a|r|} + P^2(r)~,
\end{equation}
is the analog of non-relativistic potential. Because this potential contains Dirac delta function, it becomes infinitely large at the point $r=0$. This means that if we interpret $\psi(r)$ as a wave function, it must vanish at the point $r=0$. Profile of the $V(r)$ is shown on the Figure 7.


\begin{figure}[ht]
\begin{center}
\includegraphics[width=0.7\textwidth]{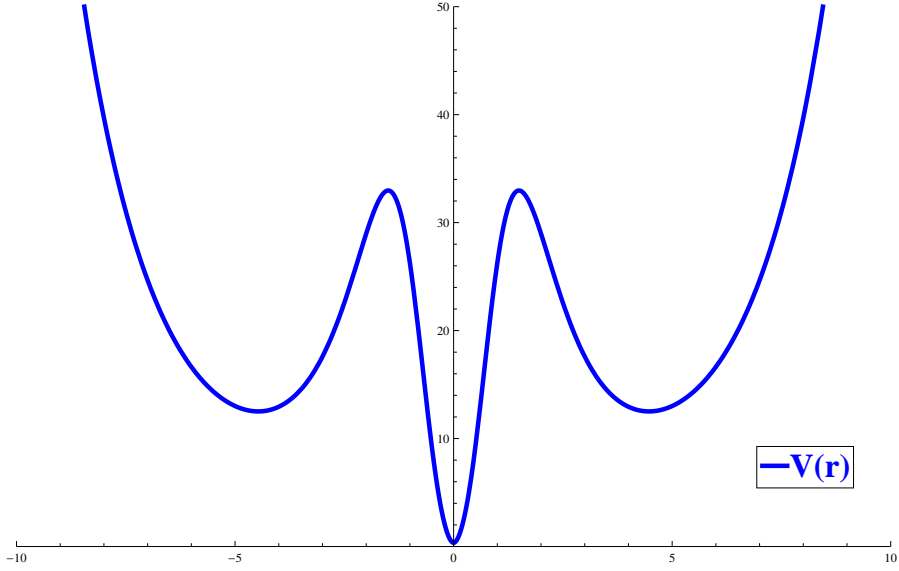}
\caption{The effective bulk potential (\ref{U}).}
\end{center}
\end{figure}


To show existence of mass gap between zero and massive modes we approximate $V(r)$ by the rectangular potential and study quantum mechanical problem in potential well with the parameters:
\begin{equation} \label{V}
V(r) = \left \{ \begin{array} {lr}
4a^2 & (0<r\leq L)\\
+\infty & (r>L)
\end{array} \right..
\end{equation}
Here $L$ is the width of the rectangular effective potential $V(r)$, which can be estimated from (\ref{U}).

Inside of the potential well the equation (\ref{psi}) obtains the form:
\begin{equation}\label{eq-psi-1}
\psi_1 '' + e^{-2a|r|} \left[ m^2- 4a^2 \right]\psi_1 = 0~.
\end{equation}
Thus in the first region, if
\begin{equation}
m^2 > 4a^2~,
\end{equation}
the solution can be written in the form:
\begin{equation} \label{phi-1}
\psi_1 (r)= C_1 J_0 \left(e^{-a|r|}\beta \right)+ C_2 Y_0\left(e^{-a|r|}\beta\right)~,~~~~~(0\leq r\leq L)
\end{equation}
where $C_1$, $C_2$ and $J_0$, $Y_0$ are integration constants and Bessel functions, respectively, and we have introduced the parameter:
\begin{equation}
\beta = \frac 1a \sqrt{m^2-4a^2}~.
\end{equation}

In the second region, where the potential is infinitely large, we have:
\begin{equation}
\psi_2(r)=0~.~~~~~(r > L)
\end{equation}

The boundary and continuity conditions at the point of junction:
\begin{equation}
\psi_1 (0)=0~, ~~~~~\psi_1 (L) = \psi_2 (L)~, ~~~~~\psi'_1 (L) = \psi'_2 (L)~,
\end{equation}
as usual, lead to the quantization of the mass:
\begin{equation}
\frac{J_0(\beta)}{J_0(e^{-aL} \beta)}=\frac{Y_0(\beta)}{Y_0(e^{-aL} \beta)}~,
\end{equation}
which can be satisfied if
\begin{equation}
\beta = \frac 1a \sqrt{m^2-4a^2} = 0~.
\end{equation}
So we find that in our model Kaluza-Klein excitations have discreet spectrum and the mass of the lightest KK massive mode is of order of the curvature scalar,
\begin{equation}
m \approx 2a~.
\end{equation}
This miens that KK states are very heavy since $a$ correspond to the energy scale characterizing the brane as a topological defect in higher-dimensional space-time.


\section{Conclusions}

In this letter within the 5D standing wave braneworld model we numerically find profiles of massless matter field wave-functions in the bulk. Using numerical solution in action integrals we have demonstrated the existences of gravitational localization of all kinds of matter fields on the brane. For the case of fermions we found that, while left ones are localized on the brane, the right massless fermions are absent on the brane, they are localized in the bulk. We have also estimated masses of KK modes and found that, as in the standard single brane models, they are very heavy - of order of the energy scale associated with the brane.


\section*{Acknowledgments}

This research was supported by the grant of Shota Rustaveli National Science Foundation $\#{\rm DI}/8/6-100/12$.


\end{document}